\documentclass{cernyrep}
\usepackage{texnames}
\usepackage[T1]{fontenc}
\usepackage{varwidth}
\usepackage{xcolor}

\begin{document}

\title{Beam Loss in Linacs}
\author{M.A. Plum}
\institute{Oak Ridge National Laboratory, Oak Ridge, Tennessee, USA}
\maketitle 

 \newcommand{\Hm}{H$^-$ }
 \newcommand{\Hn}{H$^0$ }
 \newcommand{\Hp}{H$^+$ }

\begin{abstract}
Beam loss is a critical issue in high-intensity accelerators, and much effort is expended during both the design and operation phases to minimize the loss and to keep it to manageable levels. As new accelerators become ever more powerful, beam loss becomes even more critical. Linacs for \Hm ion beams, such as the one at the Oak Ridge Spallation Neutron Source, have many more loss mechanisms compared to \Hp (proton) linacs, such as the one being designed for the European Spallation Neutron Source. Interesting \Hm beam loss mechanisms include residual gas stripping, \Hp capture and acceleration, field stripping, black-body radiation and the recently discovered intra-beam stripping mechanism. Beam halo formation, and ion source or RF turn on/off transients, are examples of beam loss mechanisms that are common for both \Hp and \Hm accelerators.
Machine protection systems play an important role in limiting the beam loss.

\end{abstract}

\section{Introduction}

Beam loss is a critical issue in high-intensity linacs, and much work is done during both the design and operation phases to keep the loss down to manageable levels. A generally accepted rule of thumb is to keep the loss to less than approximately 1 W/m to allow for hands-on maintenance. For example, the linac output beam power of the Oak Ridge Spallation Neutron Source (SNS) linac is about 1 MW today, and we plan to increase the power to its design value of 1.4 MW over the next few years and then later to  about 3~MW.  The fractional loss per metre should then be less than about $3 \times 10^{-7}$ in the high-energy portion of the linac. The allowable fraction of beam loss will be even lower for the next-generation accelerators, such as the European Spallation Neutron Source.

In general, beam loss in \Hm linacs is more difficult to manage than in \Hp linacs due to the greater number of loss mechanisms, including residual gas stripping, \Hp capture and acceleration, field stripping, black-body radiation and the recently discovered intra-beam stripping (IBSt) mechanism \cite{Shishlo2012}. Mechanisms such as beam halo formation, and ion source or RF turn on/off transients, can cause loss in both \Hp and \Hm linacs.

In this paper we will review beam loss mechanisms in \Hp and \Hm accelerators, drawing mainly from recent work at the SNS, but also including examples from other high-intensity accelerator facilities including the Los Alamos Neutron Scattering Center (LANSCE) facility at Los Alamos National Laboratory, the  ISIS facility at Rutherford Appleton Laboratory, the High Intensity Proton Accelerator (HIPA) facility at the Paul Scherrer Institute and the Japanese Proton Accelerator Research Complex (J-PARC).

\section{Why accelerate \Hm  beams?}
If \Hm  beams have so many more loss mechanisms, why should we bother with them? The reason is low-loss injection into storage rings and synchrotrons. Certain applications  require multiple beam pulses to be injected into the same RF bucket over multiple turns of injection to obtain a large beam charge per pulse. Example applications include spallation neutron sources and neutrino production facilities. The only way to inject multiple beams with low beam loss is to use charge exchange injection, where electrons are stripped off the incoming \Hm ion beam to merge it with the previously injected beam. Charge exchange injection is also required if you want the output beam emittance to be less than the sum of the input emittances (Liouville theorem).

The typical beam loss without charge exchange injection is several percent. This can be acceptable at low-power accelerators, but, for example, at SNS, where the design linac beam power is 1.4 MW, if we lose 2\% that would correspond to 28 kW, which is clearly unacceptable.
With charge exchange injection the fractional loss is just (1--2) $\times 10^{-4}$, so the beam power lost is only 140--240 W.

\section{Beam loss mechanisms}

The beam loss mechanisms that we will be discussing are listed in Table~\ref{tab:blssmech}.

\begin{table}
\caption{Beam loss mechanisms for \Hm  and \Hp  beams}
\label{tab:blssmech}
\centering
\begin{tabular}{ p{0.6\textwidth}  p{0.15\textwidth} p{0.15\textwidth} }\hline\hline
\textbf{Beam loss mechanism} & \textbf{Possible for}   & \textbf{Possible for} \\
 & \textbf{\Hm beam} & \textbf{\Hp beam} \\ \hline
Residual gas stripping  & Yes & No  \\
\Hp capture and acceleration & Yes & No  \\
Intra-beam stripping  & Yes & No   \\
Field stripping & Yes & No  \\
Black-body radiation	& Yes & No  \\
Beam halo/tails (resonances, collective effects, mismatch, etc) & Yes & Yes  \\
RF and/or ion source turn on/off transients & Yes & Yes  \\
Dark current from ion source & Yes & Yes  \\ \hline\hline
\end{tabular}
\end{table}

\subsection{Residual gas stripping}
\label{sec:RGS}

\begin{figure}
\begin{center}
\includegraphics{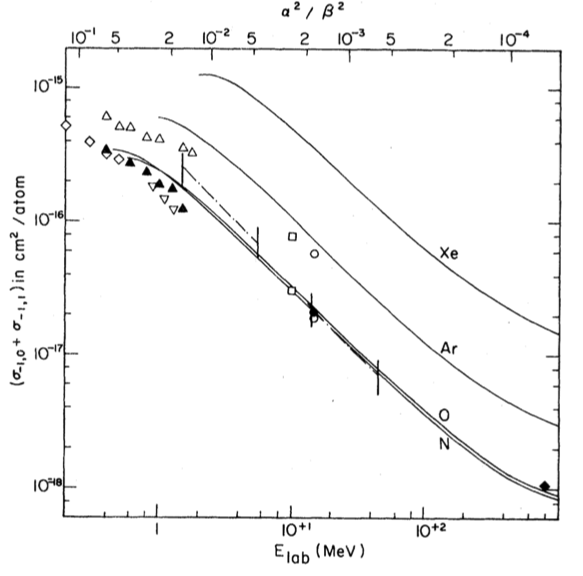}
\caption{\label{fig:Gill}  Gas stripping cross-section as a function of \Hm beam energy, for various residual gases. Figure reproduced from Ref. \cite{Gi77b}.}
\end{center}
\end{figure}

\begin{figure}
\begin{center}
\includegraphics{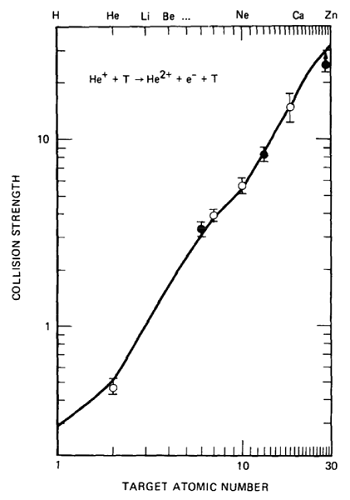}
\caption{\label{fig:Gill2}  Gas stripping cross-section as a function of atomic number. Figure reproduced from Ref. \cite{Gi84}}
\end{center}
\end{figure}

Residual gas stripping is a significant beam loss mechanism for high-power \Hm ion accelerators but not for proton accelerators.
In this mechanism electrons are stripped off the \Hm particles by the residual gas in the accelerator, most likely leaving neutral \Hn particles. The cross-sections are greatest at low beam energies and for gases with high atomic numbers, as shown in Figs. \ref{fig:Gill} and \ref{fig:Gill2}. The cross-section for double stripping (\Hm to \Hp) is about 4\% of the cross-section for single stripping (\Hm to \Hn).
The good news is that in a typical accelerator, the residual gases will be mainly H${_2}$ and H$_2$O, and then CO and CO$_2$.
These molecules have relatively low atomic numbers.

\begin{figure}
\begin{center}
\includegraphics [width=0.95\textwidth]{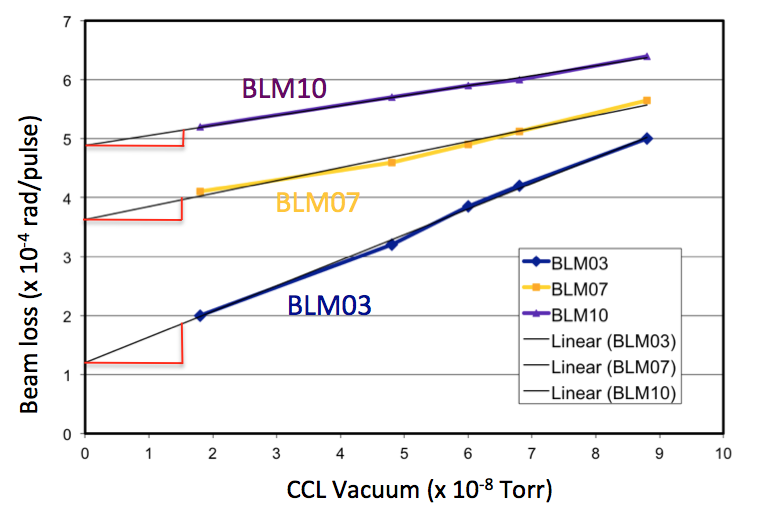}
\caption{\label{fig:RGStrip} Beam loss in the SNS SCL as a function of gas pressure in the last CCL section. The nominal gas pressure is about $2 \times 10^{-8}$ Torr, and the red brackets indicate the beam loss due to residual gas stripping during nominal operations. As one moves along the linac toward higher BLM numbers, the slopes of the lines decrease indicating less sensitivity to the upstream phenomenon, due to the BLMs being further away from the location of the loss. The magnitudes of the BLM signals also increase due to the higher beam energies. Figure reproduced from Refs. \cite{GalambosPC,Plum2012}.}
\end{center}
\end{figure}

This phenomenon is well known and understood, yet it still sometimes requires installation of additional vacuum pumps beyond those specified in the design phase. The vacuum levels can easily be worse than anticipated due to small vacuum leaks, higher-than-expected outgassing rates, etc.
In the SNS linac we have measured beam loss due to residual gas stripping in the 87 to 186 MeV coupled cavity linac (CCL) by purposely turning off the vacuum pumps and allowing the gas pressure to rise.
As shown in Fig. \ref{fig:RGStrip}, gas stripping is present to a small degree during normal operations, and it can become significant if there are vacuum problems.
There is also unexpected beam loss due to gas stripping in a section of the high-energy beam transport (HEBT) line between the linac and the ring.
Additional ion pumps were installed to mitigate this loss.
In the J-PARC linac gas stripping was found to cause significant beam loss during the commissioning phase \cite{Miura2010}.
It was subsequently reduced to acceptable levels by installing additional vacuum pumps in the Separated Drift Tube Linac (S-DTL) and the upstream portion of the linac reserved for future expansion. Also, in the LANSCE linac, residual gas stripping has been estimated \cite{Ryb2012} to cause about 25\% of the \Hm beam loss along the linac. In the ISIS linac, gas stripping is present under nominal conditions, but not at a significant level \cite{Letch2012}. However, if the gas pressure increases due to vacuum issues, the ISIS loss can become significant.

\subsubsection{Gas stripping example}

As an example of how to estimate the amount of beam loss given the reaction cross-section, consider the example of a 100 MeV, 1 mA \Hm beam, in a beam pipe with $10^{-7}$ Torr of nitrogen gas at 303 K.  The following equations are based on Ref. \cite{ref:Shafer99}.

For our beam energy range of interest, the gas stripping cross-section for nitrogen and oxygen, as shown in Fig.~\ref{fig:Gill}, may be written
\begin{equation}
\sigma = \frac{7 \times 10^{-19} }{\beta^2} \frac{\text{cm}^2}{\text{atom}}.
\end{equation}
For later reference, the cross-section for hydrogen for our beam energy range of interest may be written
\begin{equation}
\sigma = \frac{1 \times 10^{-19}}{\beta^2} \frac{\text{cm}^2}{\text{atom}}.
\end{equation}
From the ideal gas law, the gas density is
\begin{equation}
\rho = \left( 2 N_{\rm A} \frac{p}{22410 \times 760}  \frac{273}{T}    \right) \frac{\text{atoms}}{\text{cm}^3},
\end{equation}
where we note that nitrogen is a diatomic gas, $T$ is the gas temperature in K, $N_{\rm A}$ is Avogadro's constant and the pressure $p$ is in Torr.  The beam power lost in a given length of beam line $l$ is
\begin{equation}
P = E_{\rm beam} I_{\rm beam} \frac{{\rm d} \sigma}{{\rm d} \Omega} \rho l,
\end{equation}
where $E_{\rm beam}$ is the beam energy and $I_{\rm beam}$ is the beam current. Inserting the numbers, we have
\begin{equation}
P = (10^8 \Unit{}{V}) (0.001 \Unit{}{A}) \left( \frac{7 \times 10^{-19}}{0.428^2} \Unit{}{cm}^2 \right) \left(2 \times 6.022 \times 10^{23}
\frac{10^{-7} \text{ Torr}}{22410 \times 760 \text{ Torr}}
\frac{\text{ atoms}}{\text{ cm}^3}  \right)
\left( \frac{273 \text{ K}}{303 \text{ K}} \right) \left( 1 \text{ m} \right),
\end{equation}
i.e.
\begin{equation}
P = 0.243 \text{ W in 1 m or } 0.243 \text{ W/m}.
\end{equation}

\begin{figure}
\begin{center}
\includegraphics [width=0.8\textwidth]{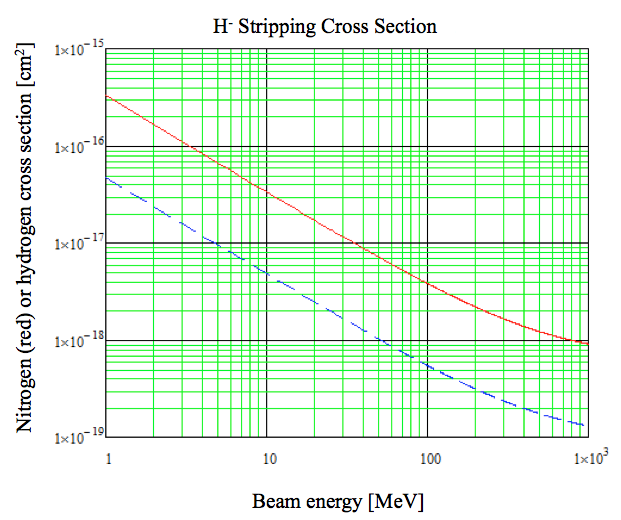}
\caption{\label{fig:CSvsE} Gas stripping cross-sections for nitrogen or oxygen (solid red line) and hydrogen (blue dashed line) as a function of beam energy.}
\end{center}
\end{figure}

\begin{figure}
\begin{center}
\includegraphics [width=0.8\textwidth]{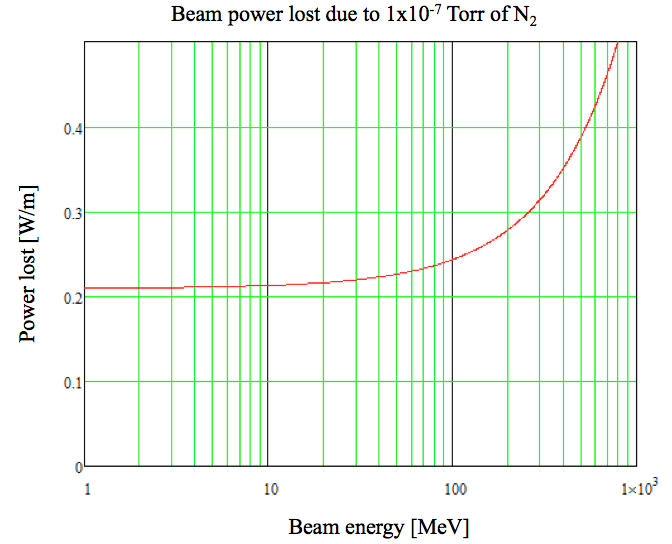}
\caption{\label{fig:PlossvsE} Beam power lost versus beam energy for our example case}
\end{center}
\end{figure}

\begin{figure}
\begin{center}
\includegraphics [width=0.8\textwidth]{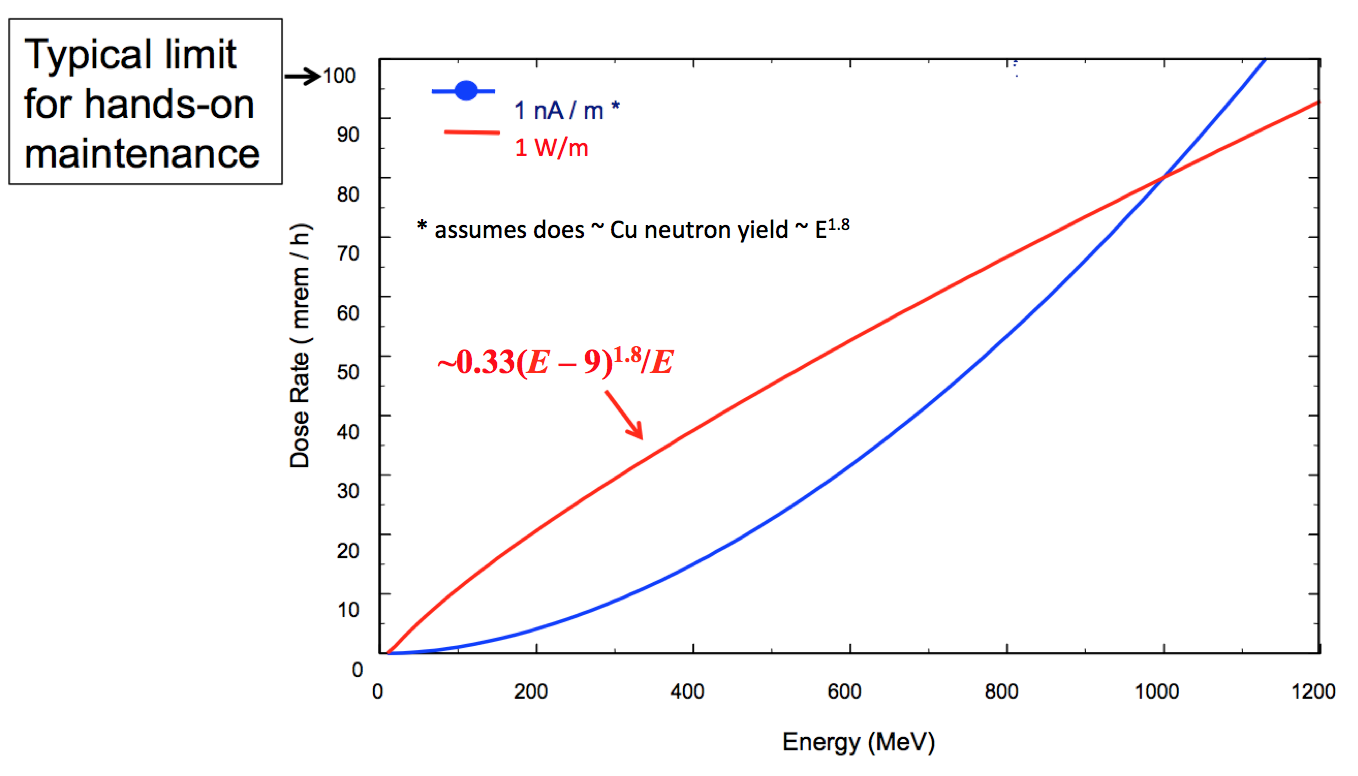}
\caption{\label{fig:dosevsE} Dose rate versus beam energy, for proton beam loss of 1 W/m or 1 nA/m, for the case of copper, at 30~cm after 4~h cool down. Figure reproduced from Ref. \cite{ref:Ga2001}. Note than 100~mrem/h = 1~mSv/h.}
\end{center}
\end{figure}

\begin{figure}
\begin{center}
\includegraphics[width=0.8\textwidth]{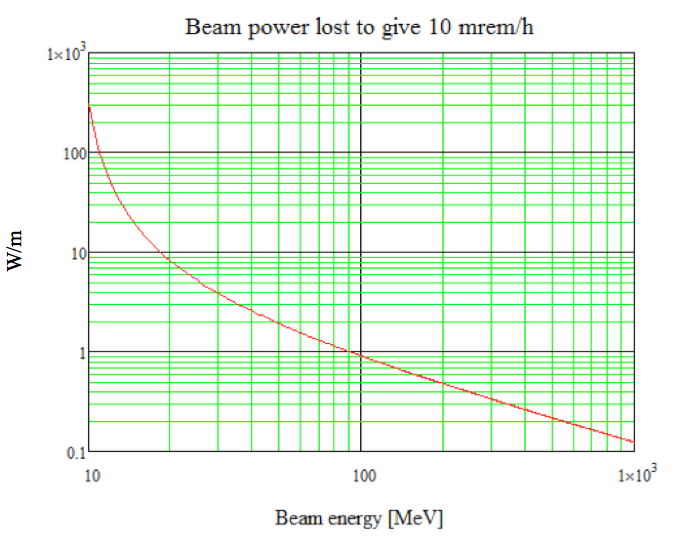}
\caption{\label{fig:Ploss10mremvsE}Allowable beam power loss versus beam energy to produce an activation of 0.1~mSv/h (10 mrem/h) at 30~cm for the case of copper, after 4~h cool down.}
\end{center}
\end{figure}

\begin{figure}
\begin{center}
\includegraphics[width=0.8\textwidth]{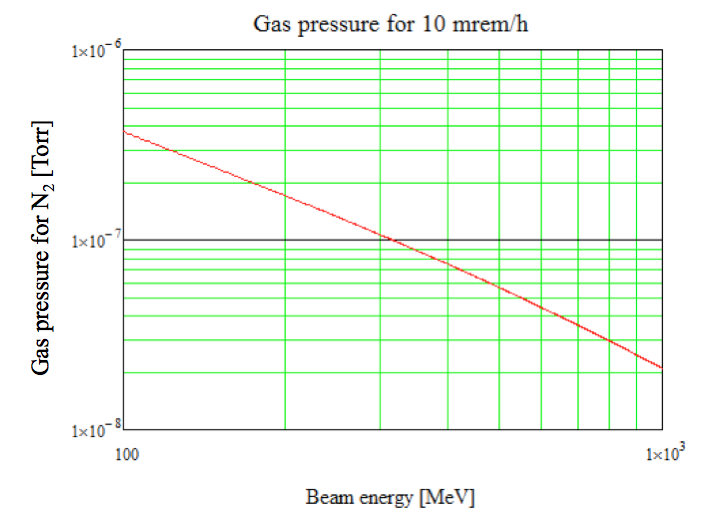}
\caption{\label{fig:GasPvsE}Allowable nitrogen gas pressure versus beam energy to produce an activation of 0.1~mSv/h (10 mrem/h) at 30~cm for the case of copper and a 1~mA \Hm beam current, after 4~h cool down.}
\end{center}
\end{figure}

Figures \ref{fig:CSvsE} and \ref{fig:PlossvsE} show the cross-sections and power lost versus the beam energy for our example case of 1 mA of a \Hm beam and $10^{-7}$ Torr of nitrogen gas at 303 K.
Note that for a given gas pressure, residual gas stripping causes the beam power lost to increase as the beam energy increases.
It is a trade-off between the cross-section decreasing with energy and the beam power increasing with energy.
The beam power wins. The next question is `is this beam loss enough to be concerned about?'. To answer that question, we need a relationship between beam loss and radioactivation. Figure \ref{fig:dosevsE} shows one example, in this case for beam loss on copper.
Copper is a good example because non-superconducting RF cavities are usually made of copper (e.g. drift tube linacs (DTLs), CCLs, superconducting linacs (SCLs), etc).
Given this relationship, we can now produce the plots shown in Figs. \ref{fig:Ploss10mremvsE} and \ref{fig:GasPvsE}.
A typical activation limit for hands-on maintenance is 1~mSv/h (100~mrem/h), from all sources.
Let us assume that we want to limit the dose from residual gas stripping to 0.1~mSv/h (10~mrem/h).
 From Fig.~\ref{fig:Ploss10mremvsE}, we see that we must limit the beam loss to 8.5~W/m at 20~MeV, 0.9~W/m at 100 MeV and 0.12~W/m at 1000~MeV, assuming the residual gas is nitrogen. Note that the allowable beam loss decreases with beam energy.
We can also look at this problem from the point of view of maximum allowable gas pressure.
 From Fig.~\ref{fig:GasPvsE}, we see that the maximum gas pressure is $1.7 \times 10^{-7}$~Torr at 200~MeV or  $2.1 \times 10^{-8}$~Torr at 1000~MeV, assuming nitrogen gas, a copper beam line and a 1~mA beam current.

\subsection{\Hp capture and acceleration}

Inadvertent proton acceleration is another interesting source of beam loss in \Hm accelerators.
As discussed in Section \ref{sec:RGS}, double stripping, where both electrons on the \Hm particle are stripped off, can occur due to residual gas interactions. If these newly created protons are captured into RF buckets, they are accelerated 180 degrees out of phase along with the \Hm particles, and then eventually lost at high beam energies.
The most likely place for \Hp capture to occur is at low beam energies, where the double-stripping cross-sections are maximized. The cross-section for double stripping is about 4\% of the single-stripping cross-section shown in Fig. \ref{fig:Gill}.
For example, recent measurements at LANSCE \cite{McC2010} showed that fully accelerated 800~MeV protons can be easily detected downstream of the linac while only the \Hm ion source is in use. The protons are from double stripping of the \Hm beam in both the 0.75~MeV low-energy beam transport (LEBT) and in the 100--800 MeV CCL.
This beam loss mechanism is also observed at J-PARC, where unexpectedly high activation levels were discovered in the beam transport line from the linac to the rapid cycling synchrotron \cite{Has2010}. Adding a chicane bump in the 3 MeV medium-energy beam transport (MEBT) solved the problem by allowing the protons to be intercepted before they could be accelerated to higher beam energies.

\begin{figure}
\begin{center}
\includegraphics{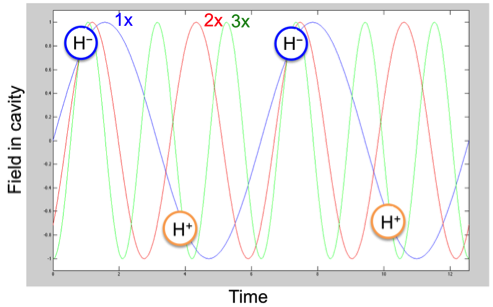}
\caption{\label{fig:HpCapture} Electric fields in a linac for three different frequencies. \Hp beams will survive a $3\times$ frequency jump since the $3\times$ phase is the same as the $1\times$ phase. This is not the case for a $2\times$ frequency jump. Figure reproduced from Ref. \cite{Plum2012}.}
\end{center}
\end{figure}

One interesting aspect of \Hp capture and acceleration is that the protons are unlikely to survive even RF frequency jumps in the linac, as illustrated in Fig. \ref{fig:HpCapture}. For example, at the Oak Ridge SNS, where the RF frequency jumps from 402.5 MHz to 805 MHz in the transition from the DTL to the CCL, the protons are unlikely to survive because they are suddenly within decelerating RF buckets after the frequency jump. Alternatively, in the J-PARC linac, where the frequency jumps from 324 MHz to 972~MHz in the transition from the S-DTL to the Annular Coupled Structure (ACS) linac at 191 MeV, the protons can be accelerated all the way to the end of the linac, only to be lost in the arc leading to the rapid cycling synchrotron.

\subsection{Intra-beam stripping}
\label{IBSt}

In IBSt \cite{Lebedev2010,Shishlo2012},  interactions between the \Hm particles within the beam bunch cause loosely bound electrons to be stripped off, leaving neutral \Hn particles, which are subsequently lost due to lack of focusing, steering and acceleration.  The IBSt reaction rate can be written as
\begin{equation} \label{eq:Leb10}
\frac{{\rm d}N}{{\rm d}s} = \frac{N^2 \sigma_{\rm max} \sqrt{\gamma^2 \theta_x^2 + \gamma^2 \theta_y^2 + \theta_s^2}}{8 \pi^2 \sigma_x \sigma_y \sigma_s \gamma^2} F( \gamma \theta_x, \gamma \theta_y, \theta_s) ,
\end{equation}
\begin{equation}
F(a,b,c) \approx 1 + \frac{2-\sqrt{3}}{\sqrt{3}(\sqrt{3}-1)} (\frac{a+b+c}{\sqrt{a^2 + b^2 + c^2}}-1),
\end{equation}
where $N$ is the number of particles in the bunch, $\gamma$ is the relativistic factor, $\sigma_{x,y} = \sqrt{ \epsilon_{x,y} \beta_{x,y}}$ are the transverse rms bunch sizes,
$\theta_{x,y} = \sqrt{ \epsilon_{x,y} / \beta_{x,y}}$ are the transverse local rms angular spreads, $\epsilon_{x,y}$ and $\beta_{x,y}$ are the transverse emittance and Twiss parameters, and
$\sigma_s$ and $\theta_s$ are the rms bunch length and the relative rms momentum spread.
$F$ is weakly dependent on its variables and ranges between 1 and 1.15.
This equation assumes a constant cross-section $\sigma_{\rm max} \approx 4 \times 10^{-15}$~cm$^{2}$ independent of the relative particle velocity, and this is correct for relative velocities $2 \times 10^{-4} < \beta_{\rm rel} < 4 \times 10^{-3} $. The cross-section drops off for $\beta_{\rm rel}$ outside this range.

During the design phase of the Oak Ridge SNS, which accelerates \Hm particles to 1~GeV with a design beam power of 1.4 MW, it was believed that the beam loss in the SCL would be negligible, due to the large apertures and low residual gas pressure. Yet, as we discovered during the commissioning phase, the beam loss was much higher than expected, with a measured fractional loss per metre of about $3 \times 10^{-7}$. Although this level of beam loss is acceptable for hands-on maintenance, it was nevertheless a puzzle.

\begin{figure}
\begin{center}
\includegraphics[width=0.62\textwidth]{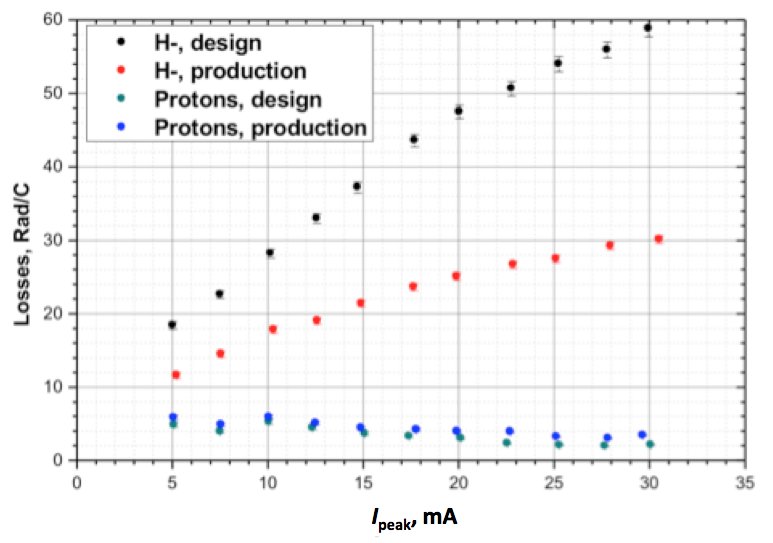}
\caption{\label{fig:Hm_Hp_lossvscurrent} Normalized beam loss (loss monitor signal divided by the peak beam current) in the SNS SCL for two different optics cases, as a function of ion source current, for both \Hp and \Hm beams. Black: \Hm beam with SCL quadrupole gradients set to design values. Green: \Hp beam with SCL quadrupole gradients set to design values. Red: \Hm beam with SCL quadrupole gradients lowered by up to 40\% to minimize the beam loss. Blue: \Hp beam with SCL quadrupole gradients set to the same values as for the \Hm minimum loss case. Figure reproduced from Ref. \cite{Shishlo2012b}.}
\end{center}
\end{figure}

The IBSt reaction rate is proportional to the beam particle density squared, which explains why, before we understood the IBSt loss mechanism at SNS, we were able to empirically reduce the beam loss by lowering the SCL quadrupole focusing strengths by up to about 40\%.
This is illustrated in Fig. \ref{fig:Hm_Hp_lossvscurrent}, which shows the normalized beam loss as a function of peak beam current, for two different cases: the SCL quadrupoles set for the design gradients and the SCL quadruple gradients lowered to empirically minimize the beam loss. The lower focusing strengths increase the transverse beam size, which lowers the beam particle density, and in turn lowers the IBSt reaction rate.
A demonstration that the IBSt reaction rate depends on the beam density squared (see Eq.~(\ref{eq:Leb10})) can also be seen in Fig. \ref{fig:Hm_Hp_lossvscurrent}, with the linear relationship between the normalized beam loss and the peak current (or quadratic relationship between the beam density and the un-normalized beam loss rate).

\begin{figure}[htbp]
\begin{center}
\includegraphics{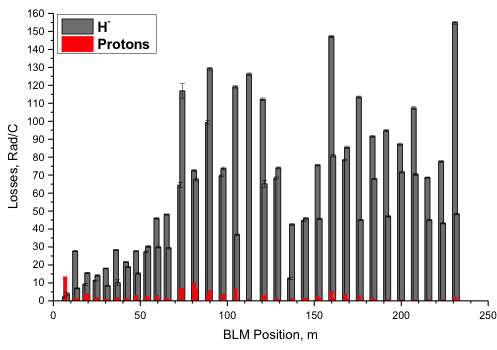}
\caption{\label{fig:HmvsHp} BLMs along the SCL, showing the proton (red) versus \Hm (grey) beam loss for the design optics case, for 30 mA beam current. Figure reproduced from Ref. \cite{Shishlo2012}.}
\end{center}
\end{figure}

Experiments accelerating protons, rather than \Hm particles, showed that IBSt is by far the dominant loss mechanism in the SNS SCL, as shown in Figs. \ref{fig:Hm_Hp_lossvscurrent} and
\ref{fig:HmvsHp}.
Similar measurements at LANSCE  \cite{Ryb2012} showed that IBSt accounts for about 75\% of the \Hm beam loss, with residual gas stripping accounting for the remaining 25\%.

For the case of the SNS SCL, the relativistic $\beta$ velocities in a given plane in the beam frame, based on particle tracking simulations,  vary between $1 \times 10^{-4}$ and $7 \times 10^{-4}$, and thereby mostly satisfy the requirements for Eq.~(\ref{eq:Leb10}).
A simple beam loss calculation based on an average stripping cross-section that is on the $\sigma_{\rm max}$ plateau, and average transverse and longitudinal beam sizes from model simulations, predicts \cite{Lebedev2010}  a total particle loss in the SCL of about $3 \times 10^{-5}$.
This value is in excellent agreement with the measured \cite{Shishlo2012}  value of (2--5) $\times 10^{-5}$.
A more precise calculation awaits a detailed particle tracking simulation.

\subsection{Magnetic field stripping}

Yet another beam loss mechanism that is important for \Hm ions but not for proton accelerators is magnetic field stripping.
This is rarely a problem since the maximum allowable fields are readily calculable and usually avoidable.
Magnetic fields are Lorentz transformed to electric fields in the rest frame of the \Hm particles and, if the field is high enough, it will strip off some electrons.
The fractional loss per unit length is given \cite{Jason81} by
\begin{equation} \label{eq:Jason81}
\frac{{\rm d}f}{{\rm d}s} = \frac{B(s)}{A_{1}}  {\rm e}^{-A_{2}/\beta \gamma c B(s)} ,
\end{equation}
where ${\rm d}f/{\rm d}s$ is the fractional loss per metre, $B(s)$ is the magnetic field as a function of distance along the beam line, $A_{1} = 2.47\times 10^{-6}$ V s/m,  $A_{2} = 4.49\times 10^{9}$ V/m and $\beta$, $\gamma$ and $c$ are the usual relativistic factors.

\begin{figure}
\begin{center}
\resizebox{0.8\textwidth}{!}{\includegraphics{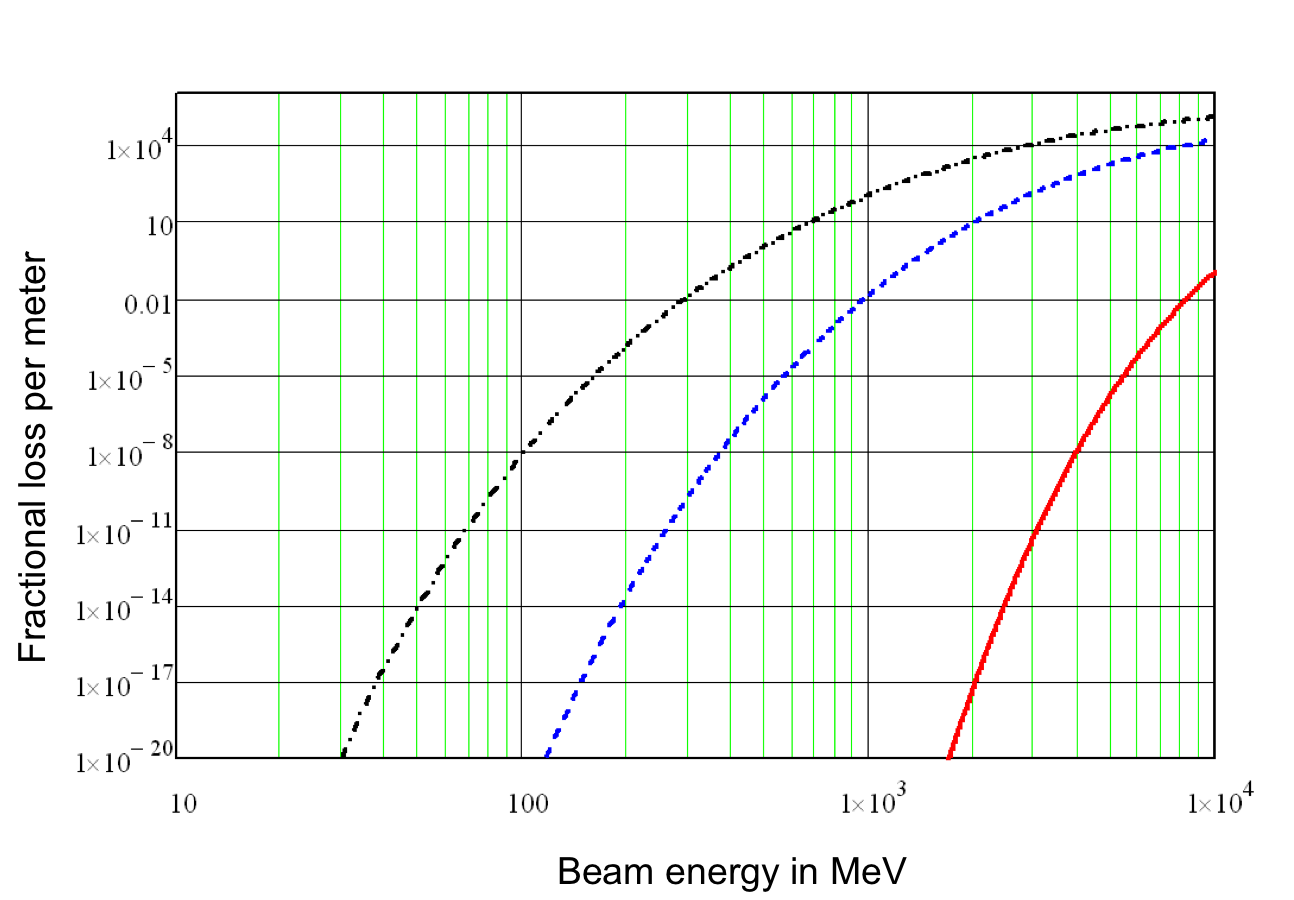}}
\caption{\label{fig:FldStrip} Fractional loss per metre due to magnetic field stripping, for magnetic fields of 0.1 (red, solid), 0.5 (blue, dashed) and 1.0 T (black, dot-dash), as a function of \Hm beam energy.}
\end{center}
\end{figure}

The effect is greatest at high beam energies where the Lorentz transform has the greatest effect, as shown in Eq. (\ref{eq:Jason81}) and plotted in Fig. \ref{fig:FldStrip}.
However, it is easy to overlook the possible scenario where, after adjusting quadrupole gradients to minimize the beam loss, the beam size is larger than expected inside quadrupole magnets whose gradients are larger than expected, which could lead to field stripping.

The ISIS facility sees a small amount of field stripping in the 70 MeV transport line between the linac and the ring, at the level of $<$1\%, just enough to create some minor hot spots \cite{Letch2012}. SNS, J-PARC and LANSCE have not reported any significant beam loss due to this mechanism.

\subsection{Black-body radiation}

Photodetachment using laser beams is a well-developed method to measure \Hm beam profiles and beam emittances, and it is now being developed as a method for charge exchange injection into storage rings and synchrotrons \cite{Liu2011}.  In these techniques, photons from a laser either strip off the loosely bound electrons from \Hm particles or excite ground-state electrons in \Hn  particles so that they can be subsequently stripped completely off by a downstream magnetic field.

Photodetachment can also be caused by black-body radiation, but the stripping rate is minimal for today's \Hm beam energies. The highest-energy \Hm beams to date are produced at the Oak Ridge Spallation Neutron Source, where the beam energy is routinely 940~MeV and the maximum beam energy, demonstrated for short times, is 1.07~GeV. At 1~GeV the beam loss rate due to room-temperature black-body radiation has been estimated \cite{Bryant2006,Johnson2008} to be just $3\times10^{-9}$ per metre or about 100 times less than our maximum allowable loss rate.

However, as the \Hm beam energy increases, the Doppler-shifted black-body photon energies can increase enough to cause significant stripping rates. For example, at 8~GeV, which is a possible charge exchange injection energy for Fermilab's Project X, the stripping rate climbs to $8\times10^{-7}$ per metre \cite{Johnson2008}. At this level of beam loss photodetachment becomes a serious concern and mitigation methods such as cooling the beam pipe to cryogenic temperatures have been considered.

The probability of beam loss due to black-body photodetachment depends on the overlap of two distributions: the \Hm photodetachment cross-section versus photon energy, which peaks at a photon energy of about 1.4~eV; and the black-body photon spectral density Doppler shifted to the rest frame of the \Hm ions. For 300 K room-temperature black-body radiation, the probability of stripping is maximum for a beam energy of about 50~GeV  \cite{Bryant2006}.

Table 2 shows a summary of the various beam loss mechanisms for some relevant  \Hm accelerators.

\begingroup
\begin{table*}[htb]
\caption{Selected beam loss mechanisms observed at various \Hm linacs}
\begin{tabular}{ p{0.17\textwidth} p{0.17\textwidth} p{0.17\textwidth} p{0.17\textwidth} p{0.17\textwidth} }
\hline \hline
\textbf{Beam loss}  & \textbf{SNS} & \textbf{J-PARC}	& \textbf{ISIS} & \textbf{LANSCE} \\
\textbf{mechanism} & & & & \\
 \hline
Intra-beam  & Yes, dominant & Not noted & Not noted  & Yes, significant,  \\
stripping & loss in SCL  & as significant  & as significant  &  75\% of loss \\
 & & & & in CCL \\
Residual gas \newline stripping & Yes, moderate \newline stripping in CCL and HEBT & Yes, significant, improved by adding pumping to S-DTL and future ACS section & Yes, not significant when vacuum is good, but can be significant if there are vacuum problems & Yes, significant, 25\% of loss in CCL \\ 
\Hp capture and acceleration & Possibly, but \newline not significant concern & Yes, was significant, cured by chicane in MEBT & Not noted as significant & Yes, significant if there is a vacuum leak in the LEBT \\ 
Field stripping & Insignificant & Insignificant & Yes, $<$1\% in 70 MeV transport line, some hot spots & 	 Insignificant \\  
Black-body \newline radiation	& Insignificant & Insignificant & Insignificant & Insignificant \\
\hline \hline
\end{tabular}
\label{default}
\end{table*}%
\endgroup

\section{Beam loss in both \Hp and \Hm linacs}

The previous sections focused on beam loss in \Hm linacs. We now turn our attention to beam loss that is common to both \Hp and \Hm linacs.
The examples we will focus on are:
\begin{itemize}
\item  beam loss due to beam halo (or beam tails) that can arise from resonances due to magnetic imperfections, collective effects (e.g. space charge) or beam mismatches;
\item  RF and/or ion source turn on/off transients;
\item dark current from the ion source.
\end{itemize}
We will discuss each of these one at a time.

\subsection{Beam loss due to halo/tails}
A well-known source of beam loss, present at the Oak Ridge SNS as well as all other proton and \Hm accelerators, is beam halo or long tails on the beam distribution. When the halo/tails intercept the beam pipe apertures the beam is lost. A certain level of halo/tail formation is inevitable, but it is exacerbated by space charge, mismatched beams, structure resonances, parametric resonances, etc.

\begin{figure}[htbp]
\begin{center}
\includegraphics [width=\textwidth]{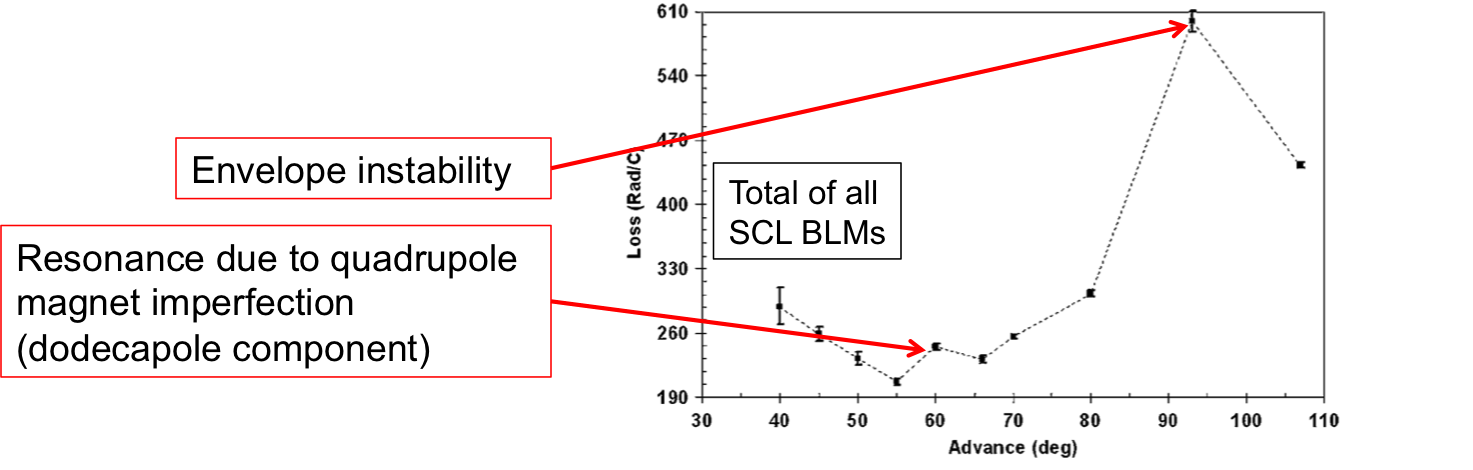}
\caption{\label{fig:resonances} Beam loss in the SNS SCL as a function of transverse phase advance per cell, with all RF cavities off.  Figure reproduced from Ref. \cite{ref:Zhang2010}.}
\end{center}
\end{figure}

Certain phase advances can cause beam loss in linacs and beam transport lines. For example, the $n\sigma_0 = 180$ or $360$~degrees resonances drive halo formation, where $\sigma_0$ is the transverse phase advance per cell for the zero space charge case.
The $\sigma_0 = 90$~degrees resonance, also known as the envelope instability, is commonly avoided in all high-intensity linac designs.
Magnet imperfections can also cause other resonance-driven losses.
An example of both of these cases, drawn from the SNS SCL, is shown in Fig.~\ref{fig:resonances}.
In this figure, the highest beam loss occurs at  $\sigma_0 = 90$~deg, and corresponds to the envelope instability.
At  $\sigma_0 = 60$~degrees we see another, much smaller, increase in beam loss due to the unwanted dodecapole component in the SNS SCL quadrupole magnets.  This is an example of a magnet imperfection resonance.

\subsection{Beam loss due to RF turn on/turn off transients}

\begin{figure}[htbp]
\begin{center}
\includegraphics [width=0.7\textwidth]{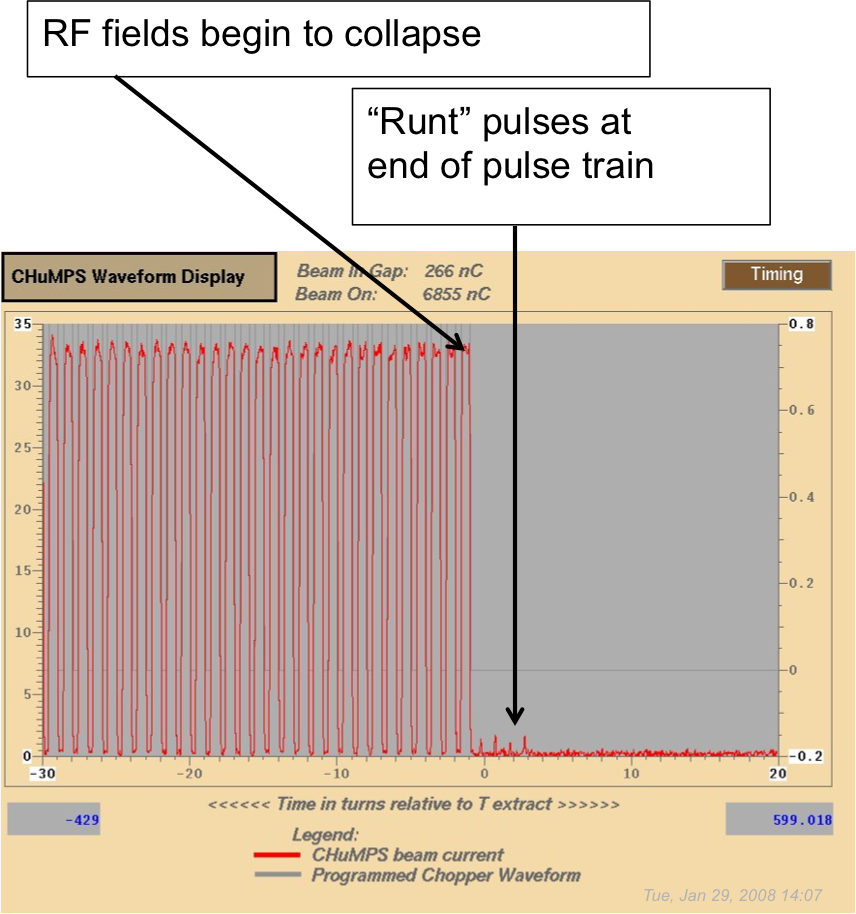}
\caption{\label{fig:runt_pulses} A fast BCM in the MEBT shows a poorly chopped beam at the end of the beam pulse \cite{Longcoy2008}. This beam will be accelerated by the downstream RF cavities, only to be lost at high energy.}
\end{center}
\end{figure}

This beam loss mechanism is important for accelerators with pulsed RF systems (as opposed to CW linacs).
If any beam  is accelerated while the RF fields are ramping up or ramping down, it is likely to be lost.
This problem is often solved with a chopper system, located at low beam energy, that blanks the beam during these times.
The SNS linac has a chopper system located just after the ion source, at a beam energy of 65 keV, but it is not perfect.
There is still a small amount of imperfectly chopped beam present at the end of the pulse.
This beam is poorly accelerated during the collapse of the RF fields, and it causes beam loss at high beam energy, primarily in the beam transport from the storage ring to the target.
We mitigate this loss by purposely ending the Radio Frequency Quadrupole (RFQ) RF pulse about 3 $\mu$s before the rest of the RF pulses.
The RFQ has a low quality factor, so its field collapses quickly, thus terminating the poorly chopped beam at low beam energy, before it can be accelerated by the rest of the linac.
The poorly chopped beam at the beginning of the pulse is mitigated by delaying the ion source turn on until the RF fields have sufficiently ramped up.
Figure \ref{fig:runt_pulses} shows an example of a poorly chopped beam at the end of the pulse. In this figure the RFQ RF pulse does not end early as discussed above.

\subsection{Beam loss due to dark current}

An unanticipated beam loss mechanism at SNS, discovered during commissioning, is dark current from the ion source. The ion source is pulsed at 60 Hz to create the required 38 mA peak \Hm ion beam, but it also emits a continuous low-level beam of about 3 $\mu$A (`dark current') due to the 13~MHz CW RF transmitter used to help ignite the pulsed plasma.
The pulsed RF accelerator systems turn on and off at 60 Hz, and the dark current is only partially accelerated during the on and off transients, which creates beam loss. Even when the dark current is properly accelerated it is undesirable, because it creates beam in the extraction gap of the storage ring.

\begin{figure}
\begin{center}
\resizebox{0.4\textwidth}{!}{\includegraphics{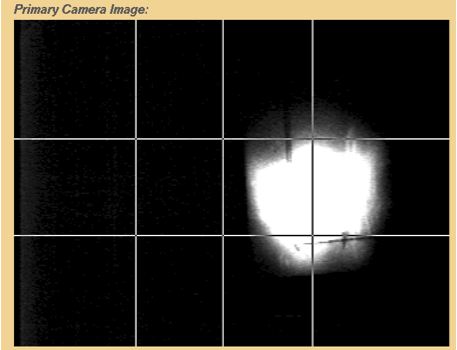}}
\caption{\label{fig:drkcurr} Example of dark current at a view screen located at the SNS ring injection point. The beam is turned off, yet the dark current is present at levels sufficient to light up the view screen. The phase of the first DTL tank is \emph{not} reversed for this image.}
\end{center}
\end{figure}

Figure \ref{fig:drkcurr} shows dark current lighting up a view screen located at the injection point of the SNS storage ring, where the stripper foil would normally be located. To mitigate this beam loss mechanism during normal operation, the LEBT chopper was modified to fully blank the head and tail of the beam pulse throughout the entire RFQ pulse length. Also, when the beam is turned off or the machine repetition rate is less than 60 Hz, the low-level RF system for the first DTL tank was modified to automatically shift the RF phase 180 degrees to prevent acceleration of the dark current beyond this point.
Both \Hp and \Hm linacs, without some kind of beam chopper system, may experience similar beam loss issues due to the ion source turn on/off transients and/or the RF turn on/off transients and/or dark current from the ion source.

\subsection{Occasional beam loss}

A large amount of beam loss can occasionally occur due to, for example:
\begin{itemize}
\item response time for RF feedback and feedforward systems;
\item RF trips off due to an interlock;
\item fluctuations in the ion source;
\item drifts in the RF system (e.g. due to temperature in klystron gallery);
\item pulsed magnets miss a pulse or provide only a partial pulse.
\end{itemize}

The integrated beam power lost may be small compared to the continuous beam loss, but the consequences can be large.
For example, occasional but large beam loss can damage superconducting cavities.
The energy deposited on the cavity surface desorbs gas or particulates, which creates an environment for arcing or low-level discharge.
This can cause the RF cavity performance to degrade over time.
At SNS, we have had to lower some cavity fields due to this damage mechanism.
Some cavities have been temporarily turned off.
The lower cavity fields cause the final beam energy to be lower.
The SNS SCL cavities do not trip off with every errant beam pulse, but the probability for a trip increases with time, and these trips cause downtime.
The cavity performance degradation from errant beam can usually be restored by warming up the cavity during a long shutdown and then RF conditioning before resuming beam operation.

RF feedback and feedforward is an important part of beam loss control.
The RF system must react to and also anticipate the beam loading caused by high-intensity beams.
Otherwise there will not be a constant accelerating field in the cavity for the duration of the beam pulse, which can cause beam loss.
Also, when the beam is turned back on after a trip, the RF system may have to re-optimize the feedback and feedforward parameters, and beam loss can be higher than normal during this time.
For example, at SNS, after a latched beam-off trip, we slowly increase the average beam current over a period of about 1 min, by ramping both the peak current and the repetition rate, to give the RF system time to adapt.
Beam losses are elevated during this time.
Also, sudden changes in the beam pulse structure can cause the beam loading to change too rapidly for the RF system to compensate, and this can also cause beam loss.
Similarly, if a RF system trips off in mid-pulse the collapsing field in the cavity will only partially accelerate the beam and cause beam loss in the downstream portion of the linac or beam transport lines.
Due to the response time of the machine protection system (15--20 $\mu$s at SNS), the ion source will continue to inject beam into the linac, only to be lost downstream of the affected cavity.

\section{Errant beam capture at SNS}

\begin{figure}
\begin{center}
\includegraphics [width=0.7\textwidth]{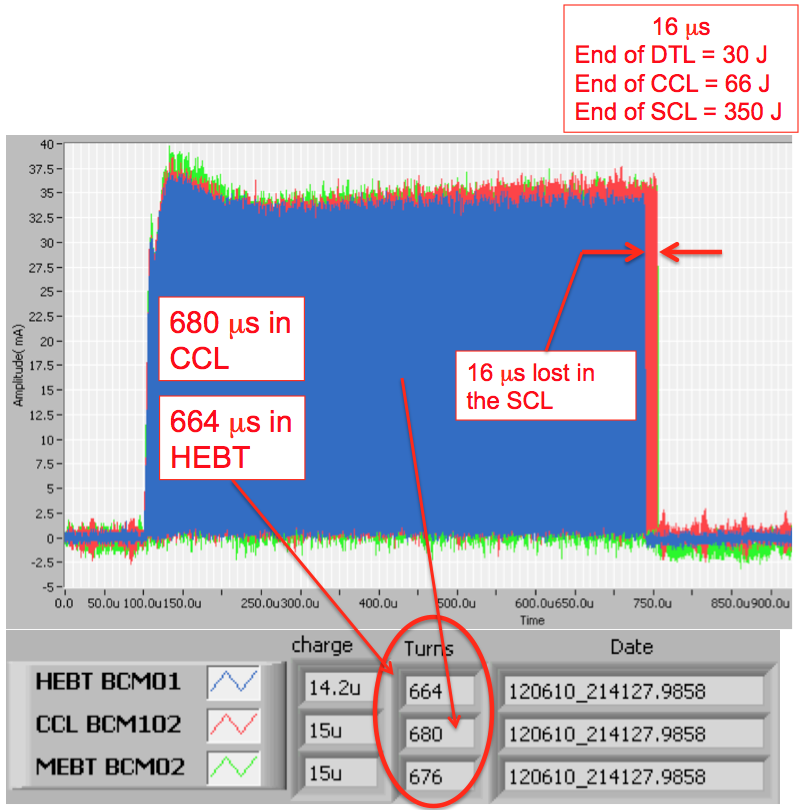}
\caption{\label{fig:errant_beam_1} Example of an errant beam event in the SNS linac \cite{Peters2013,Blokland2012}}
\end{center}
\end{figure}

\begin{figure}
\begin{center}
\includegraphics [width=0.7\textwidth]{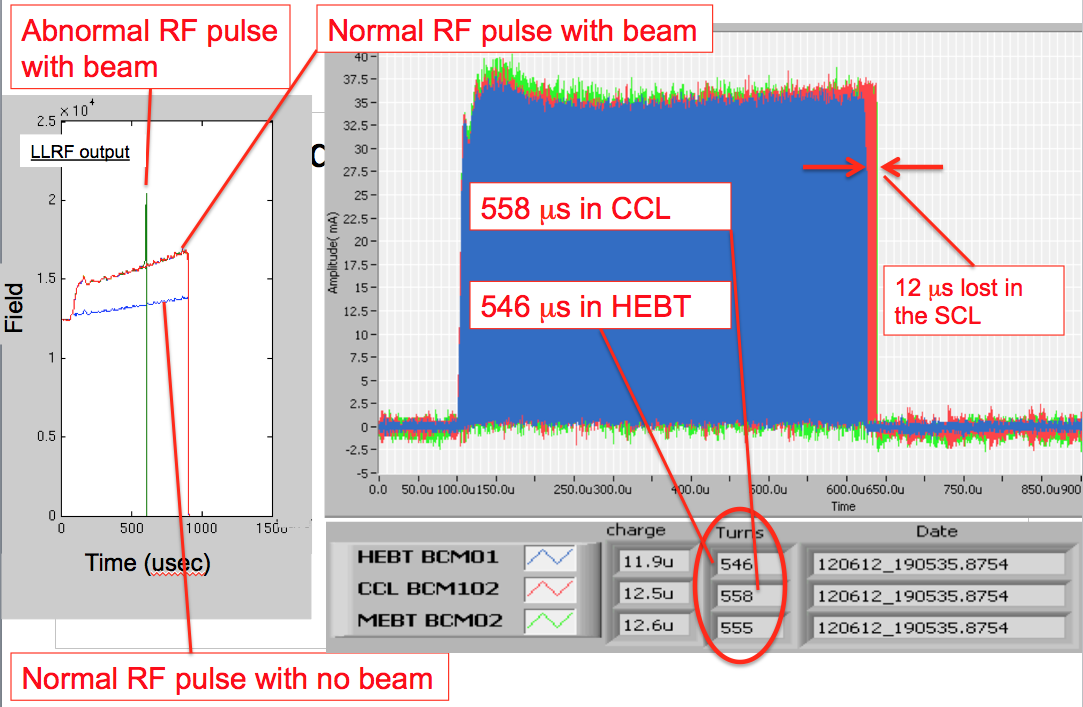}
\caption{\label{fig:errant_beam_2}  Example of an errant beam event in the SNS linac due to the sudden collapse of the RF field in one of the warm linac RF cavities \cite{Peters2013,Blokland2012}.}
\end{center}
\end{figure}

\begin{figure}
\begin{center}
\includegraphics [width=0.7\textwidth]{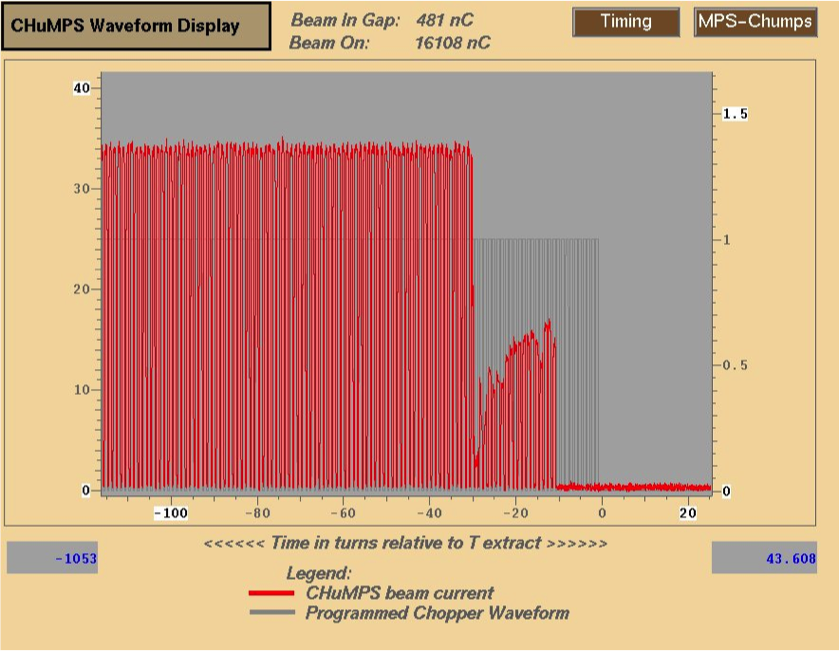}
\caption{\label{fig:errant_beam_3}  A fast BCM just downstream of the RFQ shows a sudden drop in beam current toward the end of the beam pulse \cite{Peters2013}.}
\end{center}
\end{figure}

At SNS we have assembled a system to capture signals due to errant beam events (i.e. events due to sudden occasional beam loss).
It is based on a combination of beam current monitors (BCMs), beam position monitors and beam loss monitors (BLMs).
BCMs upstream and downstream of the SCL provide a differential measurement of how much beam is lost.
The SCL BLM system, comprising 76 ion chamber detectors along the SCL, is used to identify the location of the loss.
An automated report system assembles the data for each event for offline analysis.
An example of one such analysis is shown in Fig.~\ref{fig:errant_beam_1}.
In this case the event occurred about 664 $\mu$s after the start of the pulse.
Due to the response time of the machine protection system an additional 16 $\mu$s of beam was injected into the linac, but it was lost before reaching the exit of the linac.
Based on this image alone it is not possible to say where the beam was lost and what the beam energy was, but the amount of loss corresponds to about 30 J if it is lost in the DTL, 66 J in the CCL and 350 J in the SCL.
Clearly, 350 J is enough to be concerned about for superconducting cavities.
Another example of an errant beam event in the SNS linac is shown in Fig.~\ref{fig:errant_beam_2}.
In this case about 12~$\mu$s of beam is lost in the SCL, and it is due to the sudden collapse of the RF field in one of the warm linac RF cavities.
Yet another example of an errant beam capture is shown in Fig.~\ref{fig:errant_beam_3}.
A fast BCM in the MEBT, just downstream of the RFQ, shows a drop in the peak beam current toward the end of the beam pulse, most likely due to a fault in the ion source.
The drop in peak current will cause a sudden change in beam loading in the RF cavities that will be too fast for the RF system to respond, and it will cause beam loss.

At the time of these measurements we estimated that <10\% of the errant beams were due to the ion source and the LEBT, and that most of these occurred during the first week of a new source installation, most likely due to high-voltage arcing.
More than 90\% of BLM trips were due to warm linac RF faults.
The RF faults occurred at various times during the pulse.
Those that occurred during the RF fill had reproducible times, and those that occurred during the RF flat top were at random times.
Based on these results we focused on reducing the warm linac RF faults.
We did this through a combination of empirical adjustments to the RF field amplitudes, the RF fill times, the RF resonant frequencies and by frequent cryopump reconditioning to optimize the vacuum.

\section{Beam loss mitigation}

\begin{table}
\caption{Some methods of beam loss mitigation}
\label{tab:mitigate}
\centering
\begin{tabular}{ p{0.3\textwidth} p{0.7\textwidth} }  \hline \hline
\textbf{Cause of beam loss} & \textbf{Mitigation} \\ \hline
Beam halo---both transverse and longitudinal & Scraping, collimation, better matching from one lattice to the next, magnet and RF adjustments \\  
Intra-beam stripping & Increase beam size (both transverse and longitudinal) \\  
Residual gas stripping & Improve vacuum \\  
\Hp capture and acceleration & Improve vacuum, add chicane at low energy \\  
Magnetic field stripping & Avoid by design \\  
Dark current from ion source & Deflect at low energy, reverse (phase shift) RF cavity field when beam is turned off \\  
Off-normal beams (sudden, occasional beam losses) & Turn off beam as fast as possible, track down troublesome equipment and modify to trip less often \\ \hline \hline
\end{tabular}
\end{table}

We have discussed a variety of causes of beam loss in linacs.
We will now turn our attention to methods of mitigating beam loss.
Table \ref{tab:mitigate} shows some of the more common methods and the following sections discuss some of these in more detail.

\subsection{Low-energy scraping}

A common cause of beam loss is beam tails or beam halo striking the beam pipe. The halo/tails can be present from the very beginning (at the exit of the ion source) or they can be created by mismatched beams, space charge, parametric resonances, etc. Many beam loss mitigation efforts focus on reducing the beam halo/tails. At SNS we have two main methods to control halo/tails: low-energy scraping and matching.

\begin{figure}
\begin{center}
\includegraphics[width=0.7\textwidth]{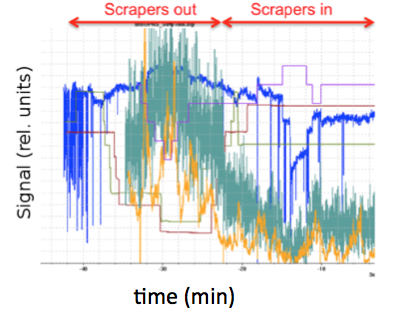}
\caption{\label{fig:Scraping} Beam loss reduction due to low-energy scraping. Magenta, light green and violet show the scraper positions. Blue shows the beam charge, and orange and dark green show BLM signals. The beam loss is reduced by up to 50--60\% at certain locations by scraping about 3\% of the beam, primarily in the  CCL, the beginning of the SCL and the ring injection dump beam line.  Figure reproduced from Ref. \cite{Galambos2010}.}
\end{center}
\end{figure}

We have found that beam scraping at low beam energy is an effective method to reduce beam loss at SNS due to halo/tails. Any scraping is best done at low beam energies where the scraper system can be more compact and where the consequent radioactivation is less.  In 2004--2005, left and right scrapers were added to the MEBT section of the warm linac, between the RFQ and the first DTL tank, where the beam energy is 2.5~MeV.
Figure \ref{fig:Scraping} shows a typical scraping result. The magnitude of the beam loss reduction varies from case to case and location to location. It is sometimes more than a factor of 10.

\begin{figure}[htb]
\begin{center}
\begin{tabular}{cc}
    \includegraphics[width=.3\textwidth]{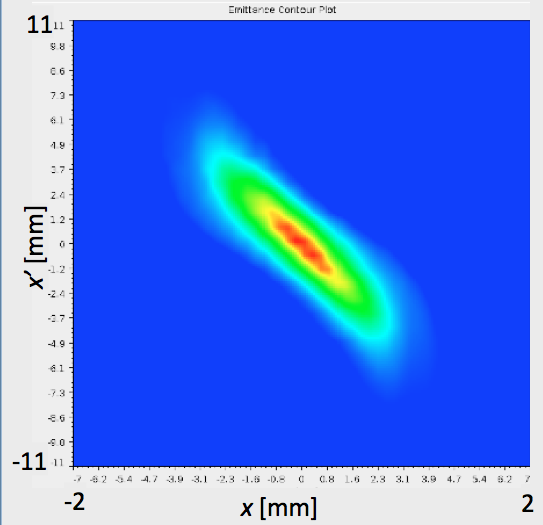} &
    \includegraphics[width=.3\textwidth]{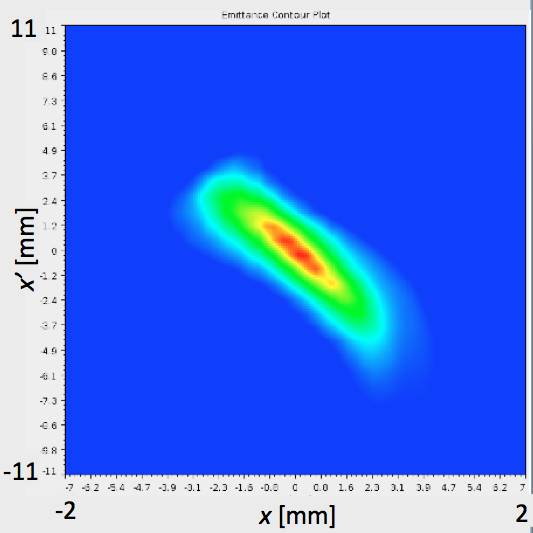} \\
\end{tabular}
\caption{\label{fig:MEBTemit} Horizontal emittance scanner without (left) and with (right) scraping. This emittance station is located in the MEBT, downstream of the scrapers, where the beam energy is still 2.5~MeV.  Figure reproduced from Ref. \cite{Zhukov2010}.}
\end{center}
\end{figure}

\begin{figure}
\begin{center}
\includegraphics[width=.6\textwidth]{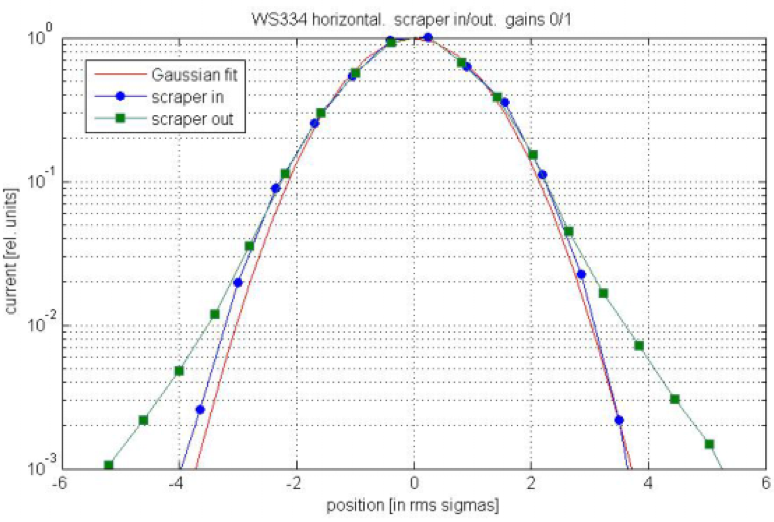}
\caption{\label{fig:ScrpDTL} A wire-scanner beam profile measurement in the DTL, before and after inserting the MEBT scrapers. The beam energy is 39.7~MeV. Blue circles---scraper is in; green squares---scraper is out; solid red line---Gaussian fit. Figure reproduced from Ref. \cite{Aleksandrov2005}.}
\end{center}
\end{figure}

\begin{figure}
\begin{center}
\includegraphics[width=.6\textwidth]{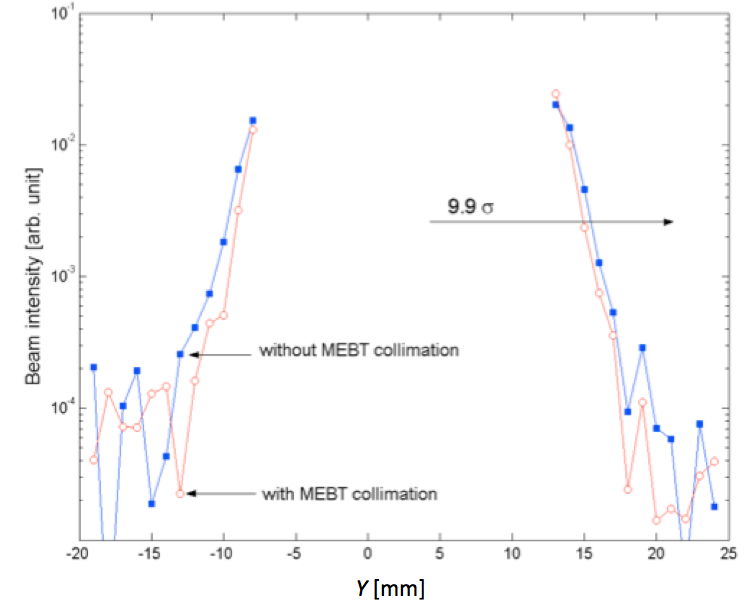}
\caption{\label{fig:ScrpHEBT} A wire-scanner beam profile measurement in the HEBT, before and after inserting the MEBT scrapers. Figure reproduced from Ref. \cite{Jeon2010}.}
\end{center}
\end{figure}

These scrapers produce a measurable reduction in the beam halo/tails, as can be seen on the MEBT slit-and-collector emittance scanner results shown in Fig. \ref{fig:MEBTemit}. The halo/tail reduction continues into the DTL, as can be seen in the non-Gaussian tails on a wire-scanner measurement shown in Fig. \ref{fig:ScrpDTL}. However, by the time the beam reaches the HEBT at the exit of the linac, there is very little measurable difference in the beam profile with and without MEBT scraping, as shown by the halo/tail measurement in Fig. \ref{fig:ScrpHEBT}. Nevertheless, as shown in Fig. \ref{fig:Scraping}, there are still large loss reductions at certain points downstream of the HEBT due to MEBT scraping, especially in the ring injection dump beam line.

Based on the success of these scrapers, we added top/bottom scrapers in summer 2013. Unfortunately there is not sufficient space in the SNS linac to add scrapers at other locations, for example between the warm and cold linac sections, but this may be desirable in the next-generation SCL designs.

\subsection{Beam loss reduction by increasing the \Hm beam size}

\begin{figure}[htb]
\begin{center}
\includegraphics[width=.52\textwidth]{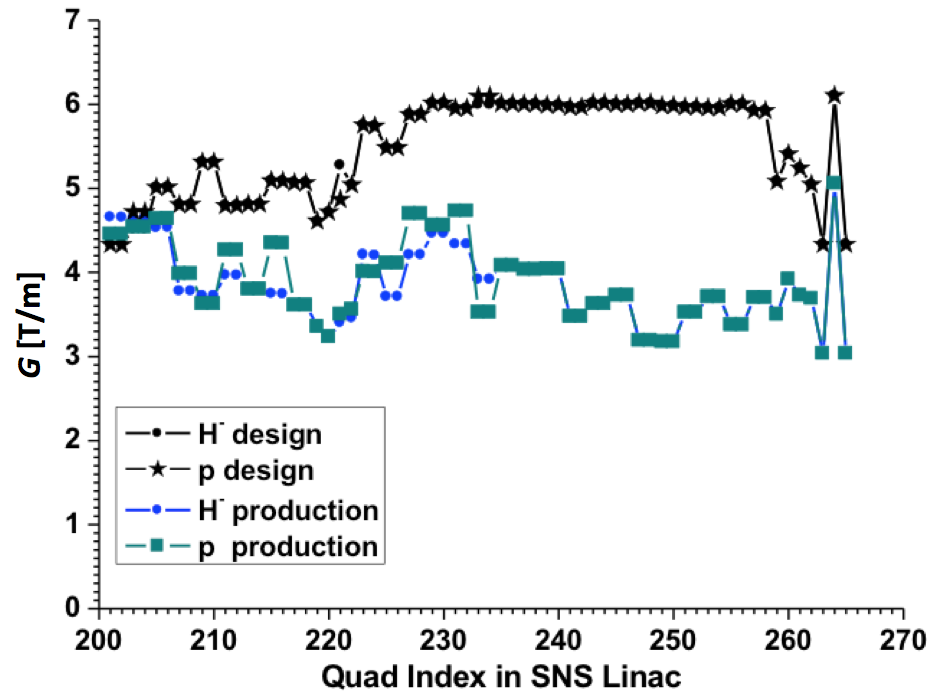}
\caption{\label{fig:scl_reduced_quads} Quadrupole strengths in the SNS SCL. Black shows the design strengths and green shows the empirically determined strengths, approximately 40\% lower than the design strengths, that reduce the beam loss by about 50\%. Figure reproduced from Ref. \cite{Shishlo2012b}.}
\end{center}
\end{figure}

In linacs that accelerate \Hm beams, IBSt can be the dominant source of beam loss.
The SNS and LANSCE linacs are two such examples.
Because the IBSt reaction rate is proportional to the beam density squared, a large reduction in beam loss can be achieved by increasing the beam size, in the longitudinal, transverse, or both, dimensions.
Figure \ref{fig:scl_reduced_quads} shows the design quadrupole strengths for the SNS SCL, and also a set of strengths that were empirically determined to reduce the beam loss by about 50\%.
The lower quadrupole strengths increase the transverse beam size, thereby lowering the beam density and lowering the beam loss, as shown in Fig.~\ref{fig:Hm_Hp_lossvscurrent}.
Further reduction in quadrupole strengths will cause the beam size to increase so much that the beam halo/tails will scrape on the beam pipe and increase the beam loss.

\subsection{Matching}

\begin{figure}[htb]
\begin{center}
\begin{tabular}{c}
    \includegraphics[width=0.6\textwidth]{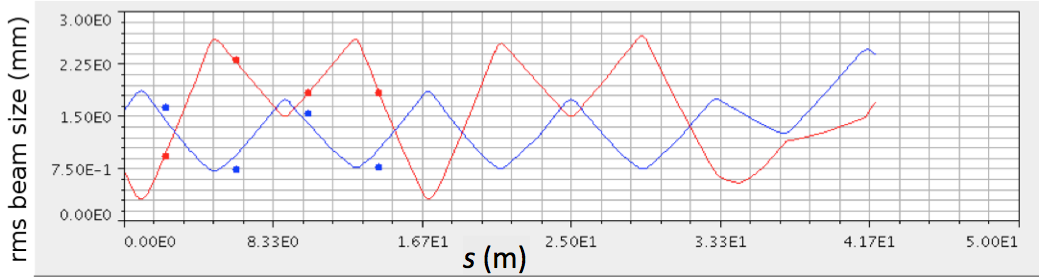} \\
    \includegraphics[width=0.6\textwidth]{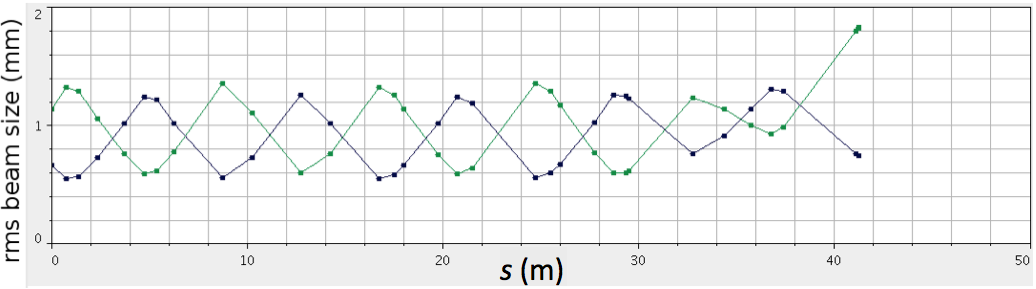} \\
\end{tabular}
\caption{\label{fig:HEBTmismatch} Beam sizes (rms) in the HEBT. Top: the low-loss production case, showing beam sizes corresponding to a mismatched beam. Dots show the measured rms beam sizes and the solid lines show the results of an envelope model adjusted to fit the data. Red is horizontal, blue is vertical. Bottom: the corresponding envelope model for the design case. In this plot the dots indicate lattice element locations. Upper plot reproduced from Ref. \cite{Cousineau2011}.}
\end{center}
\end{figure}

To minimize beam loss, conventional wisdom dictates that the Twiss parameters of the beam should be matched when the beam passes from one lattice section to the next, e.g. from one FODO lattice to another FODO lattice. For a perfect beam distribution this makes good sense because it minimizes the required aperture and prevents phase-space dilution. However, what if the beam distribution is not perfect and the Twiss parameters of the core of the beam are different from the tails of the distribution? Perhaps it is better to mismatch the core of the beam to allow better transmission (lower beam loss) for the part of the distribution that causes beam loss (i.e. the tails or halo of the beam). Figure \ref{fig:HEBTmismatch}, for example, shows measured rms beam sizes in the beginning of the SNS HEBT for the low-loss production tune, fitted with an envelope model with starting parameters varied to give the best fit. Figure \ref{fig:HEBTmismatch} also shows the design case, which assumes a well-matched beam. The low-loss tune clearly does not have a well-matched core of the beam in the horizontal plane.

\begin{figure}
\begin{center}
\resizebox{0.8\textwidth}{!}{\includegraphics{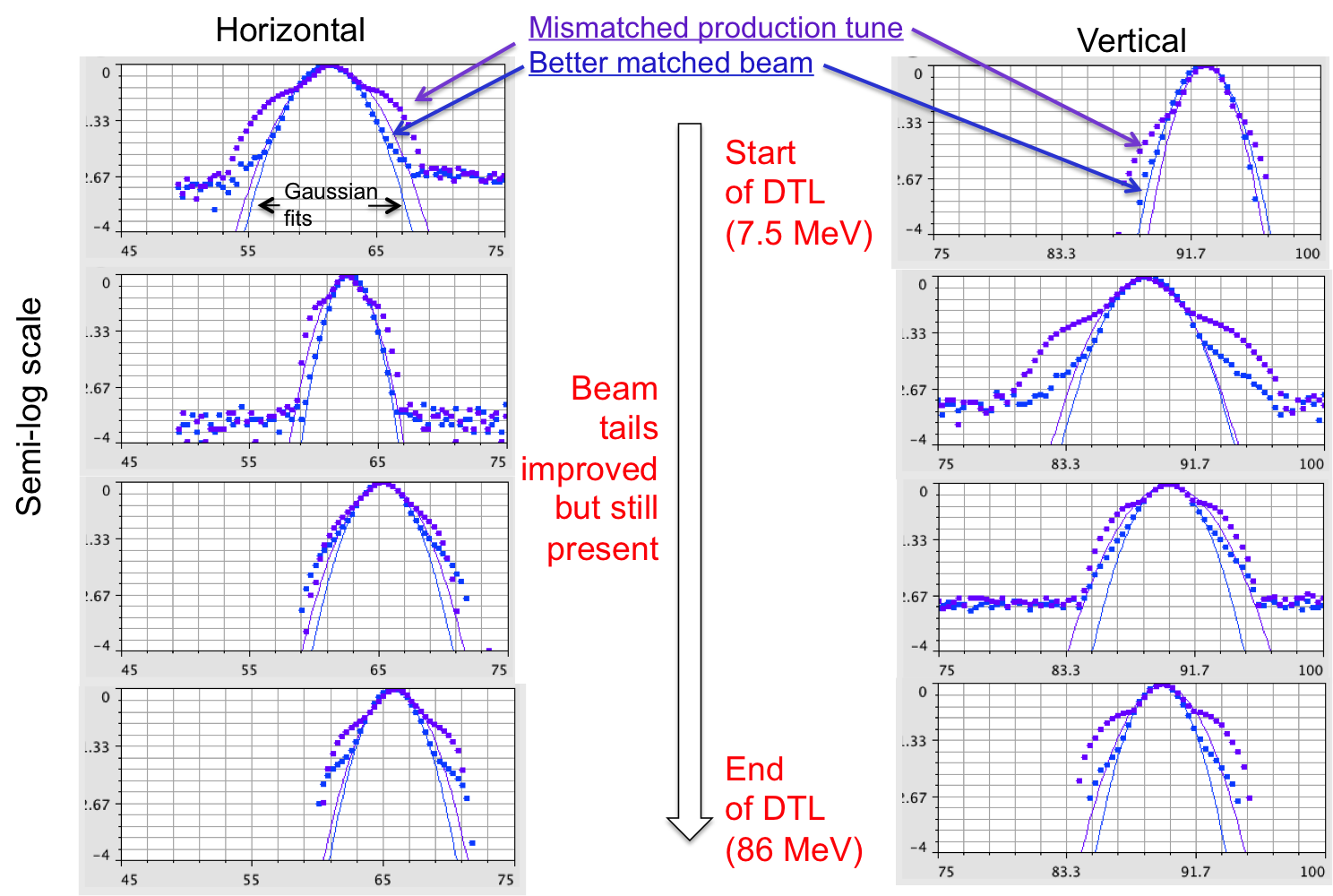}}
\caption{\label{fig:Halo} Normalized horizontal and vertical beam profiles in the SNS DTL showing beam tail formation in the DTL. Purple is the low-loss production tune and blue shows the result of matching the beam at the entrance to the DTL. The solid lines show Gaussian fits to the data.   Figure reproduced from Ref. \cite{Plum2012}.}
\end{center}
\end{figure}

Another example of beam matching versus beam tails comes from a series of large-dynamic-range wire-scanner beam profile measurements in the SNS DTL \cite{Allen2012}.
Figure \ref{fig:Halo} shows two cases: the low-loss production tune and after matching the Twiss parameters at the DTL entrance. The low-loss tune shows non-Gaussian tails as large as 30\% of the peak, starting from the first wire scanner in the DTL.
The well-matched case shows much improved non-Gaussian beam tails at the beginning of the DTL, but the losses are higher and the talks start to re-form by the end of the DTL, and then persist throughout the linac and transport lines.

At SNS, in the HEBT, the DTL and other areas of the accelerator, the low-loss tune does not have well-matched Twiss parameters for the core of the beam. Setting up the matched core case is a good place to start the beam loss minimization process, but the minimum-loss case can only be found by empirical adjustments to quadrupole gradients and the RF phases and amplitudes. With our present suite of beam instrumentation it is not possible to accurately characterize the parameters of the beam tails/halo, and then use that data to better match the beam, because of the limited dynamic ranges of the measurements.

\subsection{Beam loss reduction by magnet and RF phase adjustments}

As discussed in the previous section, when the SNS accelerator complex is operated with the quadrupole gradients set to the design values, the beam loss is high and the same is true for some of the RF cavity phases.
We have empirically found that the loss can be reduced by at least a factor of two by making small adjustments to the matching quadrupole gradients and to certain RF phases.
The minimum beam loss case is very sensitive to certain RF phases, such as those in the DTL. Just 1~degree of phase change can double the beam loss at certain locations.
As the SNS accelerator complex has matured, we have worked to bring the quadrupole gradient set points in line with design values from simulation codes.
In some cases the simulation codes have been updated to more accurately reflect the actual accelerator, and in other cases low-loss tunes have been found that minimize the discrepancies between the accelerator and the model.  The concept is that model-based adjustments that have a physics basis are better than random adjustments that may lead to a localized minimum in beam loss or to an operating point that is difficult to sustain.

The SNS DTL quadrupole gradients are by definition operated according to design, since they are permanent magnets. The CCL gradients are now also operated very close their design values, starting in August 2012. The SCL quadrupoles are up to 40\% below their design values, to increase the beam size as discussed above. The HEBT quadrupole gradients are mostly operated according to their design values, with the biggest exceptions being at the beginning and end of the HEBT, where the lattice changes from the SCL to the HEBT, and from the HEBT to the ring. The beam loss in the injection dump beam line is very sensitive to changes in the HEBT quadrupole gradients.

Similar experiences are seen at other accelerator facilities, including LANSCE \cite{Rybarcyk2012}, ISIS \cite{Adams2012} and the HIPA cyclotron at PSI. At HIPA the last 20--50\% of beam loss reduction comes from empirical tuning \cite{Seidel2012}.

\section{Summary}

Beam loss is a very important driver to the cost, design and operation of high-intensity \Hp and \Hm accelerators, and will become even more important as next-generation accelerators come online. Beam loss in \Hm accelerators is much more complex than \Hp accelerators due to the many more beam loss mechanisms available to \Hm beams. Modern design and simulation codes cannot accurately predict beam loss and so prudent designs must include flexible machine lattices and mitigation measures such as scrapers and collimators. However, the future is bright for the next-generation accelerators that can take advantage of and expand on the observations and experiences of today's high-intensity accelerator facilities.

\section*{Acknowledgements}
ORNL is managed by UT-Battelle, LLC, under contract DE-AC05-00OR22725 for the U.S. Department of Energy.

Notice: This manuscript has been authored by UT-Battelle, LLC, under Contract No. DE-AC05-00OR22725 with the U.S. Department of Energy. The United States Government retains, and the publisher, by accepting the article for publication, acknowledges that the United States Government retains a non-exclusive, paid-up, irrevocable, world-wide licence to publish or reproduce the published form of this manuscript, or allow others to do so, for United States Government purposes.


\end{document}